\UseRawInputEncoding
\documentclass[aps,prl,reprint,twocolumn,smsmath,amssymb,superscriptaddress]{revtex4-2}
\usepackage{amsfonts}
\usepackage{amsmath}
\usepackage{amssymb}
\usepackage{charter}
\usepackage{graphicx}
\usepackage{float}
\usepackage[colorlinks=true,linkcolor=blue,citecolor=blue,urlcolor=blue]{hyperref}
\usepackage[all]{hypcap}
\usepackage[defaultcolor=red]{changes}
\newcommand{\prlsection}[1]{\textit{#1.-}}

\begin{document}

\title{Beating the Bad-Cavity Limit via Auxiliary-Emitter Linewidth Squeezing}
\author{Youke Xu}
\author{Zeyang Liao}\email{liaozy7@mail.sysu.edu.cn}
\author{Xue-hua Wang}\email{wangxueh@mail.sysu.edu.cn}
\affiliation{\textit{State Key Laboratory of Optoelectronic Materials and Technologies, School of Physics, Sun Yat-sen University, Guangzhou 510275, China}}

\begin{abstract}
Strong coupling in cavity QED is conventionally achieved at the expense of either high cavity quality factors or ultrasmall mode volumes, a trade-off that fundamentally constrains practical implementations. Here, we circumvent this limitation by introducing two nonidentical auxiliary emitters with opposite detunings into a bad cavity. This hybrid system supports a subradiant mode that significantly squeeze the effective cavity linewidth, creating an ultra-narrow transmission window at the cavity frequency. As a result, a target emitter placed in this engineered environment exhibits prolonged vacuum Rabi oscillations and resolved spontaneous emission splitting, which are clear signatures of strong coupling, even though the bare cavity remains in the weak-coupling regime. Our scheme thus transforms a bad cavity into an effective platform for strong-coupling physics, with potential applications in quantum computation and quantum sensing.
\end{abstract}
\maketitle

\prlsection{Introduction}The coherent exchange of energy between photons and matter, enabled by strong light-atom coupling, forms the basis for applications ranging from quantum computing to quantum sensing \cite{Reagor2016,Zhang2026,Knaut2024,Redchenko2025,Wang2024,Wang2026,Li2026,Liao2024,XuNT2024,Cirac2025}. In free space, however, the intrinsically weak interaction makes strong coupling between light and atoms a formidable challenge.
Confining the electromagnetic field into a small mode volume via a cavity significantly enhances light-matter interaction and greatly facilitates reaching the strong coupling regime \cite{Wei2025,Yu2019,Ge2021}, making cavity quantum electrodynamics (cavity QED) a prominent platform for quantum information processing and fundamental quantum optics \cite{McKeever2003,Ohta2011,4-PhysRevResearch.6.033295,Muniain2025,XingPRA2024,6-PhysRevLett.130.216901,7-Svendsen_Kurman_Schmidt_Koppens_Kaminer_Thygesen_2021,8-PhysRevB.107.045425,Wei2024PLA}.

Vacuum Rabi splitting in emission spectra (or equivalently, vacuum Rabi oscillations in the time domain) are commonly regarded as hallmarks of strong coupling in cavity-QED systems. To clearly resolve such splitting or oscillations, the energy level splitting must exceed the linewidth, which typically demands that the light-matter coupling strength 
$g$ be comparable to or larger than the system's dissipation rate \cite{Santhosh2016,27-Khitrova2006,28-Torma2014}.
The coupling strength scales as $g \propto 1/\sqrt{V_m}$, with $V_m$ being the mode volume, while the cavity linewidth $\kappa$ is inversely proportional to the quality factor $Q$ of the cavity. Therefore, achieving strong coherent light-matter interactions requires a relatively high $Q/V_m$ ratio. 

However, in conventional cavity designs, there is usually an inherent trade‑off between the quality factor and the mode volume: a higher $Q$ typically comes with a larger mode volume \cite{Goktug-2022-OE}. For instance, dielectric cavities (e.g., whispering-gallery mode cavities) can achieve extremely high $Q$ factors, but their mode volumes are relatively large due to diffraction limit \cite{36-PhysRevLett.118.263202,37-Hunger_Steinmetz_Colombe_Deutsch_H,38-Corato-Zanarella_Gil-Molina_Ji_Shin_Mohanty_Lipson_2022,39-Toropov_Cabello_Serrano_Gutha_Rafti_Vollmer_2021,40-Zhang_Su_Liu_Xing_Sum_Xiong_2016,45-PhysRevLett.108.227402,46-Brossard_Xu_Williams_Hadjipanayi_Hugues_Hopkinson_Wang_Taylor_2010,47-Ota_Takamiya_Ohta_Takagi_Kumagai_Iwamoto_Arakawa_2018,48-Kawata:24}. In contrast, plasmonic cavities offer ultra‑small mode volumes at the expense of a low quality factor due to significant ohmic losses \cite{41-PhysRevA.85.031805,42-PhysRevLett.116.253904,43-Zhang:20,44-PhysRevApplied.18.044066}. Consequently, reducing the cavity linewidth (i.e., increasing $Q$) without sacrificing a small mode volume has become a crucial research direction, as it would enable strong coupling even with modest coupling strengths and open new possibilities for quantum information processing. 

Considerable efforts have been devoted over the past few decades to reducing the effective cavity linewidth. Prominent passive strategies include the use of intracavity electromagnetically induced transparency (EIT) \cite{Lukin1998,Gessler2007,Wu2008,Mucke2010,Lin2016}, the exploitation of topologically protected dissipationless edge states \cite{Lu2024}, control of the Purcell effect via exceptional point \cite{LuPRA-2023,Agarwal2024}, and the integration of plasmonic cavities with low-loss dielectric cavities \cite{Peng2017,Bitton2021,Lusciencechina2021,Li-prl-2023,Rathi2025}. Meanwhile, advances in inverse design have enabled the creation of topologically optimized cavities that achieve a high quality-factor-to-mode-volume ratio \cite{Wang2018,51-7frd-pf1m,52-Hu_Khater_Salas-Montiel_Kratschmer_Engelmann_Green_Weiss_2018,53-Dong_Babar_Christiansen_Hansen_Stobbe_Yu_M,Yan2025-NC,Wang2026-SA}. In contrast to these passive approaches, active manipulation of the system via parametric driving provides a versatile means to exponentially enhance the light-matter coupling strength, thereby enabling strong or even ultrastrong coupling in conventional cavity QED setups \cite{54-PhysRevLett.120.093602,55-8qtt-symt}.

In this letter, we show that introducing two nonidentical auxiliary emitters with opposite detunings into a bad cavity can significantly reduce the effective cavity linewidth, enabling strong coupling in a regime where it would otherwise be absent. The hybrid system develops an ultra-narrow transmission window at $\omega_c$—the hallmark of a subradiant mode. A target emitter coupled to this mode exhibits Rabi oscillations and a split spontaneous emission spectrum, in stark contrast to the behavior of the bare cavity.



\begin{figure}
\centering \includegraphics[width=0.95\columnwidth]{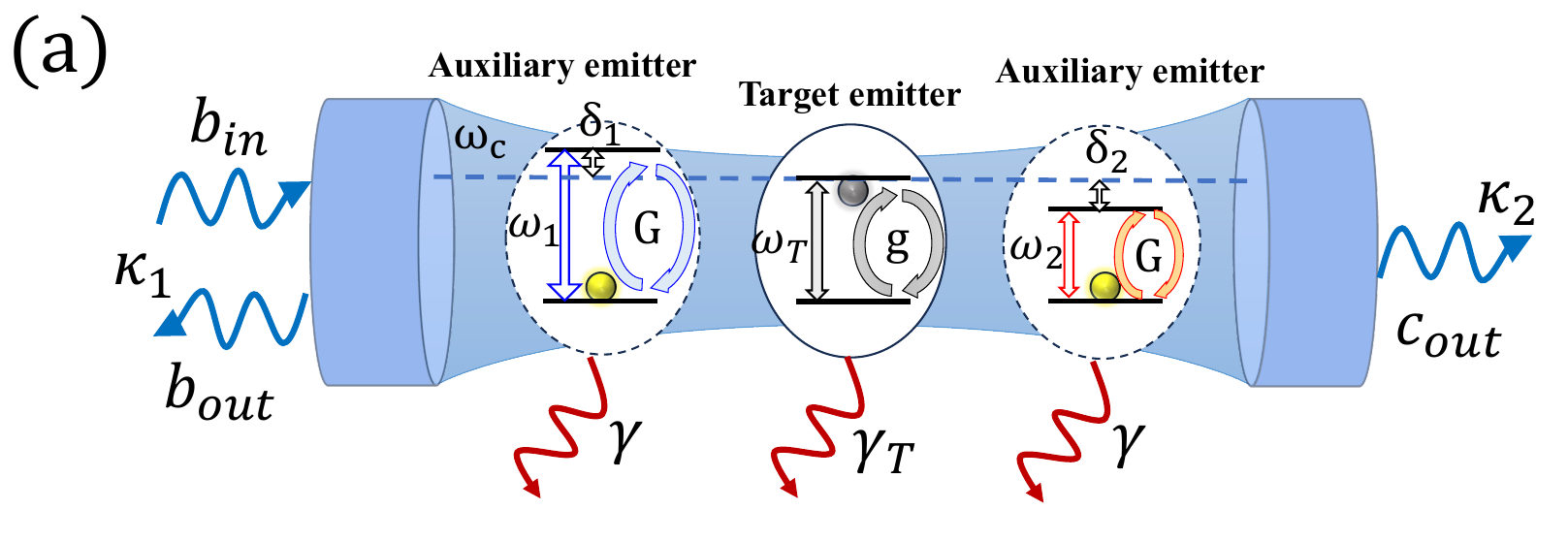}
\centering \includegraphics[width=0.95\columnwidth]{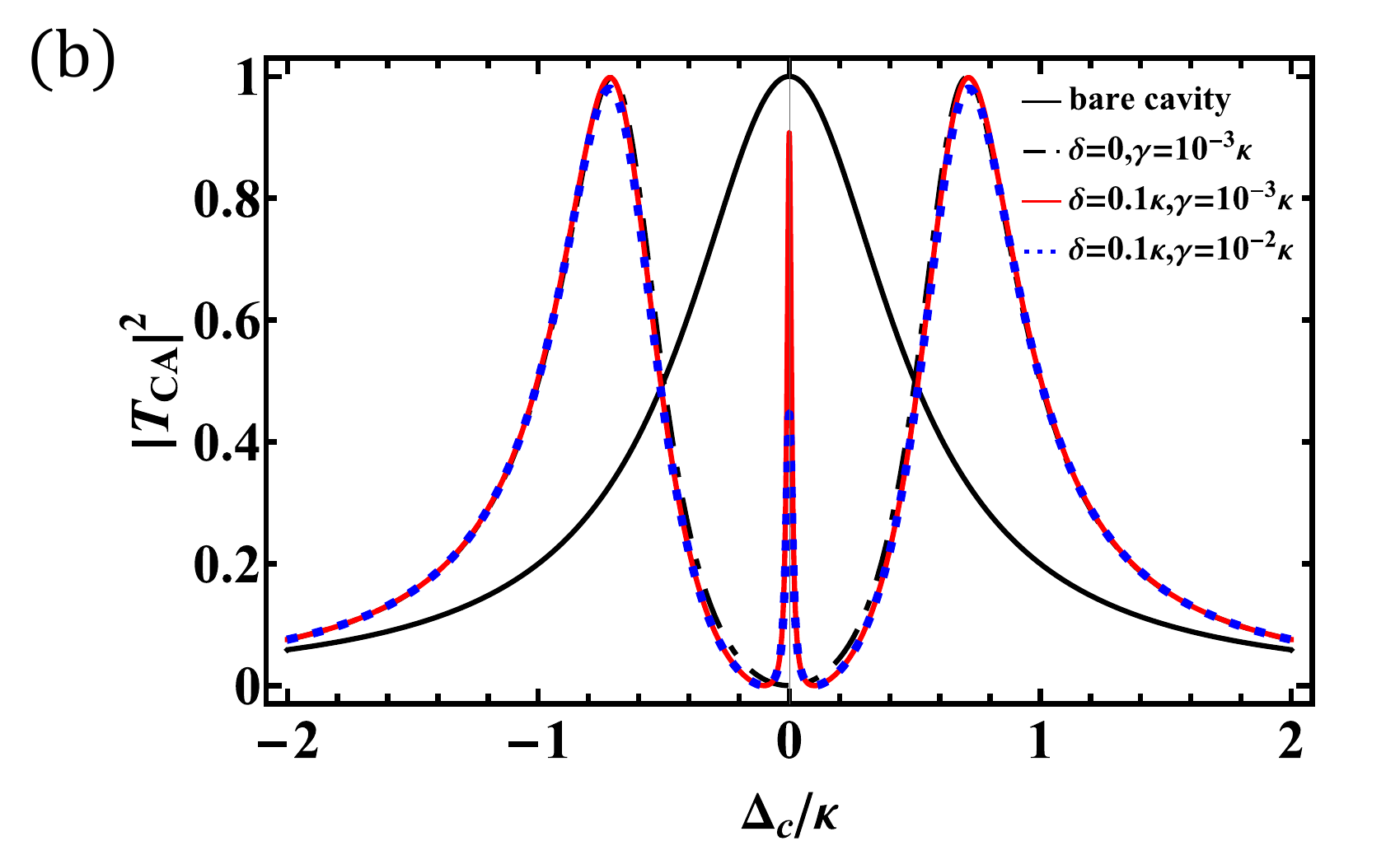}
\caption{(a) Schematic of linewidth reduction and strong coupling in a bad cavity assisted by two non-identical auxiliary emitters. (b) Transmission spectra for a bad cavity with two non-identical auxiliary emitters (detuning $\delta = 0.1\kappa$). The red solid and blue dotted lines correspond to $\gamma=0.001\kappa$ and $\gamma=0.01\kappa$, respectively. For comparison, the bare cavity and the case of two identical emitters are shown as the black solid and black dashed-dotted lines, respectively. Other parameters: $G = 0.5\kappa$. \label{Fig1}
}
\end{figure}

\prlsection{Model and Hamiltonian}
As is shown in Fig.~\hyperref[Fig1]{\ref{Fig1}(a)}, a bad cavity with frequency $\omega_{c}$ and linewidth $\kappa$ is coupled to two auxiliary non-identical emitters and a target emitter. The decay rate $\gamma$ of each auxiliary emitter is much smaller than $\kappa$, and their transition frequencies are denoted as $\omega_{1}$ and $\omega_{2}$, respectively. The target emitter possesses a transition frequency $\omega_{T}$ and a decay rate $\gamma_{T}$. Working in the rotating frame with respect to the cavity frequency $\omega_{c}$, the Hamiltonian of the coupled system is given by (setting $\hbar=1$) \cite{Wei2025}
\begin{align}
H =&\sum_{j=1}^{2}\delta_{j}\hat{\sigma}_{j}^{+}\hat{\sigma}_{j}^{-} +\sum_{j=1}^{2}\left( G\hat{\sigma}_{j}^{+} \hat{a}+H.c.\right) \nonumber \\  &+\delta_T\hat{\sigma}_{T}^{+}\hat{\sigma}_{T}^{-} +\left( g\hat{\sigma}_{T}^{+} \hat{a}+H.c.\right) \label{system Hamiltonian}
\end{align}%
where the first line corresponds to the cavity–auxiliary emitter coupling Hamiltonian $H_{CA}$, and the second line to the cavity–target emitter coupling Hamiltonian $H_{CT}$. Here, $\delta_{j}=\omega_{j}-\omega_{c}$ denotes the detuning between the $j$th auxiliary emitter and the cavity, and $\hat{\sigma}_{j}^{+}=|e_{j}\rangle\langle g_{j}|$ ($\hat{\sigma}_{j}^{-}=|g_{j}\rangle\langle e_{j}|$) represents the raising (lowering) operator of the $j$th emitter. The operators $a^{\dagger}$ and $a$ are the cavity creation and annihilation operators, respectively. The two auxiliary emitters are coupled to the cavity mode with identical strength $G$. For the target emitter, $\delta_{T}=\omega_{T}-\omega_{c}$ is its detuning from the cavity frequency, and $g$ denotes its coupling strength to the cavity.

\prlsection{Effective cavity linewidth reduction}To illustrate how the auxiliary emitters reduce the effective cavity linewidth, we first neglect the target emitter in the following analysis. The Hamiltonian is then given by the first line of Eq. (\ref{system Hamiltonian}).
From the quantum Langevin equations, we can obtain
\begin{eqnarray}
\dot{\hat{\sigma}}_{j} &=&-\left( i\delta _{j}+\frac{\gamma }{2}\right)
\hat{\sigma} _{j}-i G \hat{a},  \label{eq2} \\
\dot{\hat{a}} &=&-\frac{\kappa }{2}
\hat{a}-i\sum_{j=1}^{2}G^{\ast }\hat{\sigma} _{j}-i\sqrt{\kappa_1}\hat{b}_{in},
\label{eq3}
\end{eqnarray}%
where $\kappa = \kappa_1 + \kappa_2$ denotes the total cavity leakage rate, with $\kappa_1$ and $\kappa_2$ being the leakage rates of the left and right mirrors, respectively. Here, $%
\hat{b}_{in}\left( t\right) $ is the input field, and the input-output
relationship reads%
\begin{eqnarray}
\hat{b}_{out} &=&i\sqrt{\kappa_1}\hat{a}-\hat{b}_{in},  \label{eq4} \\
\hat{c}_{out} &=&-i\sqrt{\kappa_2}\hat{a}.  \label{eq5}
\end{eqnarray}%
For simplicity, in all subsequent discussions, we set the leakage rates at
both ends of the cavity to be equal, i.e., $\kappa _{1}=\kappa _{2}=\frac{%
\kappa }{2}.$ Based on Eqs. (\ref{eq2}-\ref{eq5}) and employing the Fourier
transform relationship $O\left( t\right) =\frac{1}{\sqrt{2\pi }}\int d\omega
e^{-i\omega t}O\left( \omega \right) ,$ the relationship between the
transmission field and the input field can be derived as $\hat{c}_{out}\left( \Delta_c \right) =T_{CA}(\Delta_c)\hat{b}_{in}\left(\Delta_c\right)$, 
where the transmission coefficient of the coupled system reads as (see supplementary material)%
\begin{equation}
T_{CA}(\Delta_c)=\frac{\frac{\kappa }{2}}{i\Delta _{c}-\frac{\kappa }{2}+\sum_{j=1}^{2}%
\frac{\left \vert G\right \vert ^{2}}{i\left( \Delta _{c}-\delta _{j}\right)
-\frac{\gamma }{2}}}.  \label{eq7}
\end{equation}%
Here, $\Delta _{c}=\omega -\omega _{c}$ is the detuning between the incident field and the cavity resonant frequency.

The transmission spectrum $|T_{CA}|^2$ as a function of the detuning $\Delta_c$ is shown in Fig.~\hyperref[Fig1]{\ref{Fig1}(b)}. For a bare cavity without emitters, the transmission spectrum exhibits a typical Lorentzian lineshape with linewidth $\kappa$ (black solid line). 
In contrast, when two auxiliary non-identical emitters ($\delta_1 = -\delta_2=\delta \neq 0$) are introduced, the transmission spectrum displays three peaks featuring an extremely narrow central peak (red solid line for $\gamma=0.001\kappa$, blue dotted line for $\gamma=0.01\kappa$, and more results are shown in the supplementary material). This observation indicates that the effective cavity linewidth is significantly reduced. The separation between the two broaden side peaks is approximately $2\sqrt{2}G$ due to collective effects. However, if the two auxiliary emitters are identical ($\delta_1 = \delta_2 = 0$), the narrow central peak vanishes, leaving only two side peaks. Equation (\ref{eq7}) shows that for $\delta_1 = \delta_2 = 0$, $T_{CA}(0) = (1 + 8|G|^2 / \gamma\kappa)^{-1}$. When $\gamma \ll |G|, \kappa$, $T_{CA}(0) \to 0$, meaning the central peak disappears. In contrast, if $\delta_1 = -\delta_2 = \delta$ with $|\delta| \gg \gamma$, the second term in the denominator cancels, yielding $T_{CA}(0) \to -1$ and thus a transparency window. Consequently, the ultra-narrow transmission window arises from destructive interference between two reflection channels, which occurs only when the two auxiliary emitters possess detunings of opposite signs \cite{Liao2016,Lei2023}.


\begin{figure}[tph]
\centering \includegraphics[width=1\columnwidth]{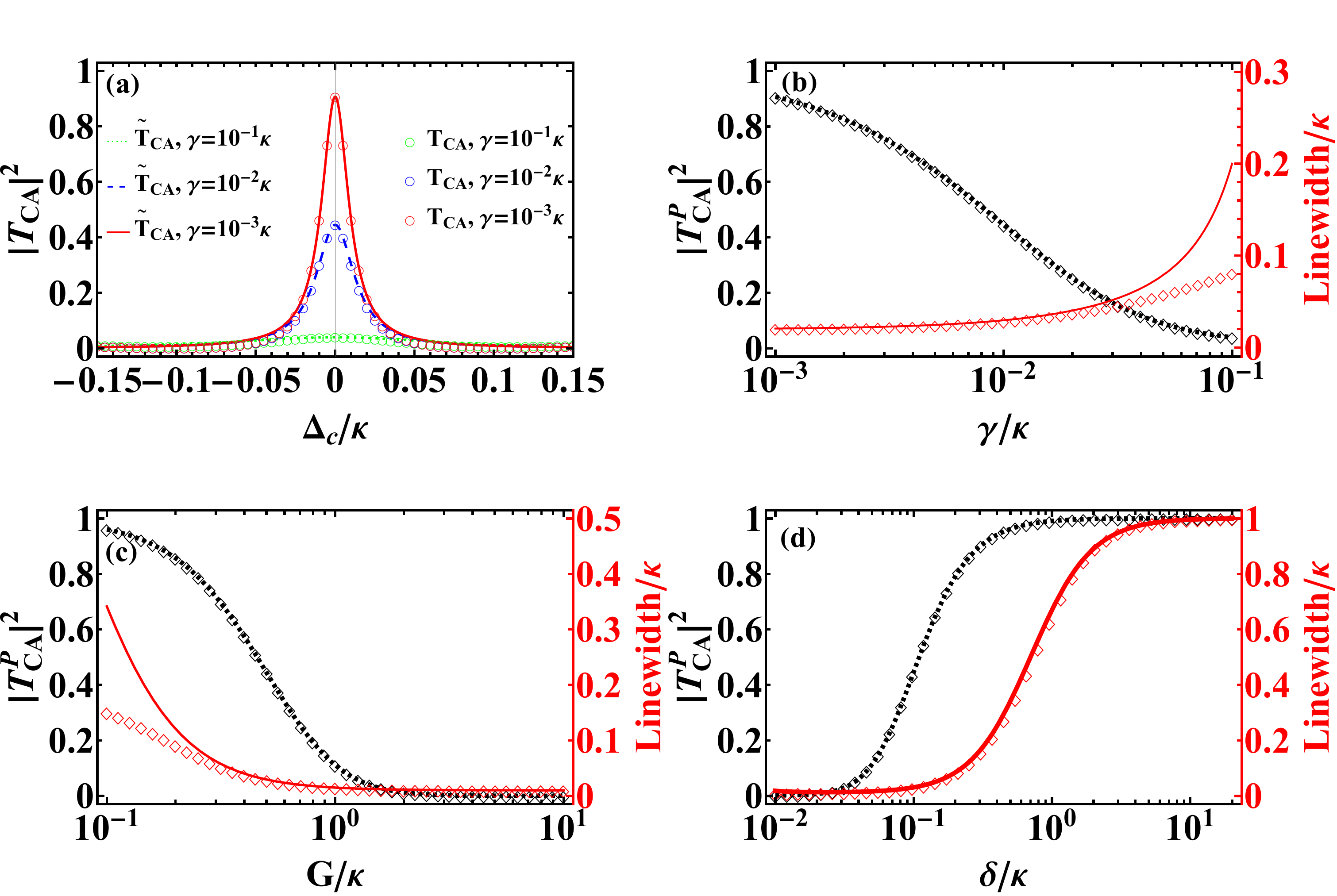}
\caption{(a) Central transmission peaks for a cavity with two non-identical emitters ($\delta_1 = -\delta_2 = \delta = 0.1\kappa$) at three different values of $\gamma$.  Empty-circles: exact results (Eq.~\ref{eq7}); colored lines: approximate results (Eq.~\ref{eq8}). (b-d) Transmission peak probability (black dashed line) and linewidth (red solid line) as functions of (b) $\gamma$, (c) $G$, and (d) $\delta$, with fixed parameters specified in each panel: (b) $G = 0.5\kappa$ and $\delta = 0.1\kappa$; (c) $\delta = 0.1\kappa$ and $\gamma = 0.01\kappa$; (d) $G = 0.5\kappa$ and $\gamma = 0.01\kappa$. The black and red diamond-shaped marks denote the maximum values of the transmission probability and linewidth, respectively, obtained from Eq.~(\ref{eq7}).
\label{Fig2}}
\end{figure}

Next, we analyze the characteristics of the central peak and their dependence to the parameters of the auxiliary-emitter cavity hybrid system, assuming $\delta_1 = -\delta_2 = \delta > \gamma$. In the vicinity of the resonant frequency, Eq. (\ref{eq7}) can be approximated as (see supplementary material)
\begin{equation}
\tilde{T}_{CA}(\Delta_{c})=\frac{\kappa\eta/2}{i\Delta_{c}-\frac{\kappa_{eff}}{2}} \label{eq8}
\end{equation}
where 
\begin{equation}
\kappa_{eff}=\eta\Big(\kappa+\frac{2|G|^2\gamma}{\Omega^2}\Big) \label{eq9}
\end{equation}
is the effective linewidth of the hybrid cavity and the parameter $\eta=\Omega^4/[\Omega^4+2G^2(\delta^2-\gamma^2/4)]$ with $\Omega^2=\delta^2+\gamma^2/4$.
Figure \hyperref[Fig2]{\ref{Fig2}(a)} shows the central transmission peak for three different values of $\gamma$. The transmission peak probability decreases with $\gamma$, while the linewidth increases. The empty circles represent the exact results from Eq.~(\ref{eq7}), and the colored lines denote the approximate results from Eq.~(\ref{eq8}). The excellent agreement between the two sets of results confirms the validity of the simplified expression.

From Eq. (\ref{eq8}), we can see that the peak transmissivity at resonant frequency ($\Delta_c=0$) is given by 
\begin{equation}
|T^{p}_{CA}|^2=\Big(1+\frac{2\gamma G^2}{\kappa\Omega^2}\Big)^{-2}. \label{eq10peak transmissivity}
\end{equation}
Under the condition that $\gamma\ll \delta, G<\kappa$, we have the effective linewidth $\kappa_{eff}\approx \eta\kappa\approx \kappa/(1+2G^2/\delta^2) $ and the peak intensity $|T^{p}_{CA}|^2\approx \Big(1+2\gamma G^2/\kappa\delta^2\Big)^{-2}$. As shown in Figs. \hyperref[Fig2]{\ref{Fig2}(b-d)}, the effective linewidth $\kappa_{\text{eff}}$ monotonically decreases with increasing $G$ and with decreasing $\delta$ and $\gamma$. In contrast, the central peak intensity $|T^{p}_{CA}|^2$ increases as $G$, $\delta$, or $\gamma$ decreases. The validity of these observations is supported by Figs. \hyperref[Fig2]{\ref{Fig2}(b-d)}, where the analytical results from Eqs.~(\ref{eq8}) and (\ref{eq9}) (solid lines) are compared with the exact numerical results from Eq.~(\ref{eq7}) (symbols). The approximate expressions are in good agreement with the exact results, except in the regime of small $G$ or large $\gamma$. In the limit $G=0$, the system reduces to a bare cavity, for which $\kappa_{\text{eff}} = \kappa$ and $|T^{p}_{CA}|^2 = 1$. In the limit $\delta \rightarrow 0$, both $\kappa_{\text{eff}}$ and $|T^{p}_{CA}|^2$ approach zero, indicating that the transmission peak vanishes when the two auxiliary emitters are identical.

\begin{figure}[tph]
\centering \includegraphics[width=1\columnwidth]{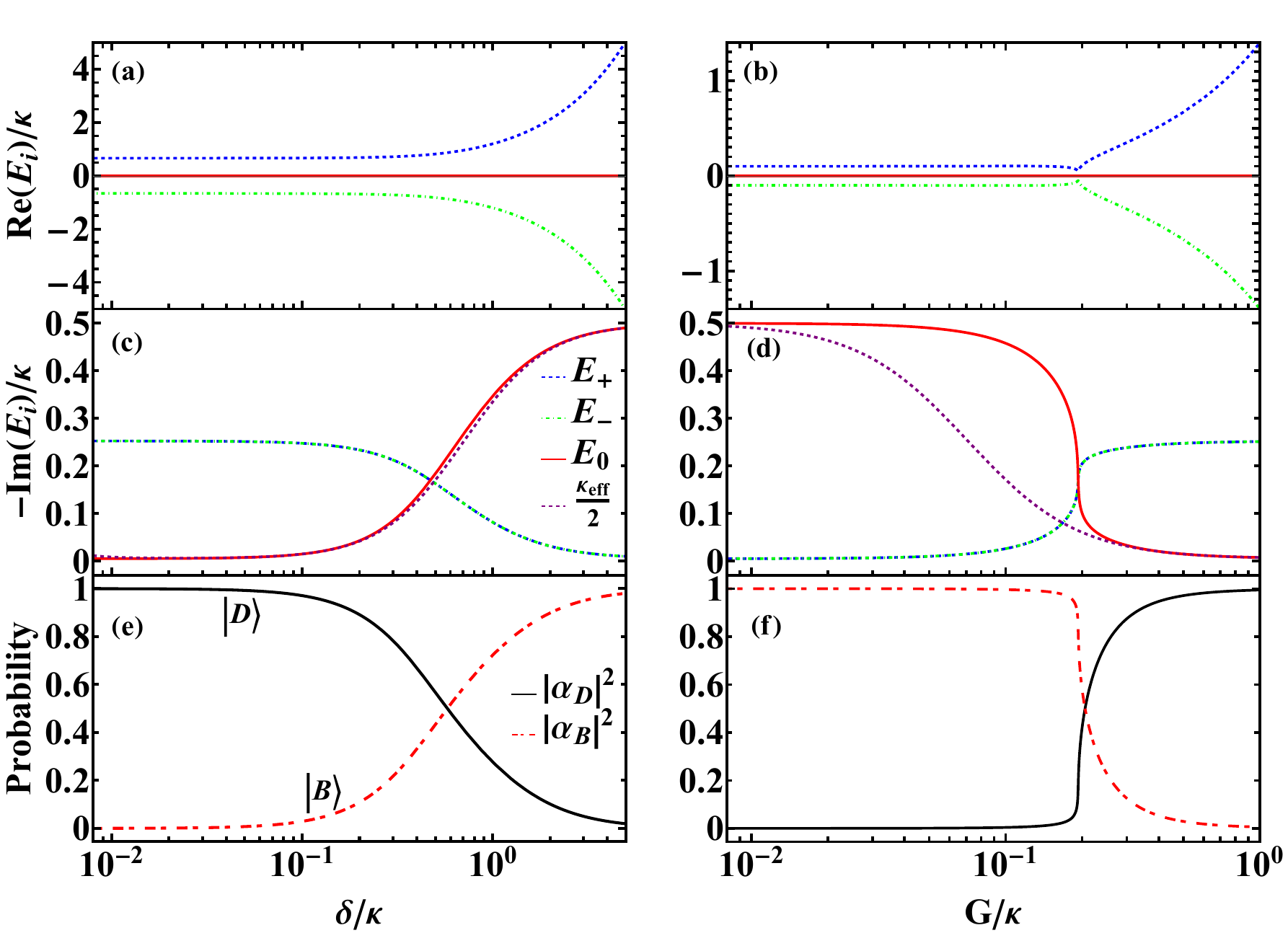}
\caption{The real (a,b) and imaginary (c,d) parts of the eigenvalues of the hybrid system for different values of $\delta$ and $G$. $E_0$ corresponds to the central peak, and the other two edge peaks correspond to $E_+$ and $E_-$, respectively. (e-f) The probabilities of the dark state component and the bright state component of the eigenstate of the central peak. Parameters: (a,c,e) $\gamma=0.01\kappa$ and $G=0.5\kappa$; (b,d,f) $\gamma=0.01\kappa$ and $\delta=0.1\kappa$.
\label{Fig3}}
\end{figure}

\prlsection{Subradiant polariton}The ultranarrow linewidth can also be explained via the eigenvalue analysis of the effective Hamiltonian:
$H_{CA}^{\text{eff}} = H_{CA} - i\frac{\gamma}{2}\sum_{j=1}^{2}\hat{\sigma}^{+}_{j}\hat{\sigma}^{-}_{j} - i\frac{\kappa}{2}\hat{a}^{\dagger}\hat{a}$.
In the single-excitation subspace, three eigenvalues emerge, one of which corresponds to the central mode ($\operatorname{Re}(E)=0$, red line in Figs. \hyperref[Fig3]{\ref{Fig3}(a)} and \hyperref[Fig3]{\ref{Fig3}(b)}). This eigenvalue is purely imaginary, with its magnitude dropping below $0.1\kappa$ for $\delta < 0.5\kappa$ or $G > 0.2\kappa$ (red line in Figs. \hyperref[Fig3]{\ref{Fig3}(c)} and \hyperref[Fig3]{\ref{Fig3}(d)}), and in this regime Eq.~(\ref{eq8}) matches $\operatorname{Im}(E_0)$ well (purple dot-dashed line). Thus, the central mode becomes highly subradiant with two nonidentical emitters. Its eigenstate is approximately $|\psi_0\rangle = \alpha_B |B\rangle + \alpha_D |D\rangle$, where $|B\rangle$ combines $(|eg0\rangle + |ge0\rangle)/\sqrt{2}$ and $|gg1\rangle$, and $|D\rangle = (|eg0\rangle - |ge0\rangle)/\sqrt{2}$ (see supplementary material). As shown in Figs. \hyperref[Fig3]{\ref{Fig3}(e)} and \hyperref[Fig3]{\ref{Fig3}(f)}, $|\alpha_B|^2$ and $|\alpha_D|^2$ versus $\delta$ and $G$ reveal that the eigenstate is mostly in the dark state for $\delta < 0.5\kappa$ and $G > 0.2\kappa$. Hence, the central mode is essentially a subradiant polariton which has very narrow linewidth.

\prlsection{Strong coupling in the hybrid cavity} In the above discussions, we have shown that by introducing two non-identical emitters in a bad cavity, the effective cavity linewidth of the hybrid cavity can be significantly reduced. Here, we further show that by placing a target emitter inside this hybrid cavity (Fig. \hyperref[Fig1]{\ref{Fig1}(a)}), weak to strong coupling transition can be achieved. 

\begin{figure}[tbh]
\centering
\includegraphics[width=1\columnwidth]{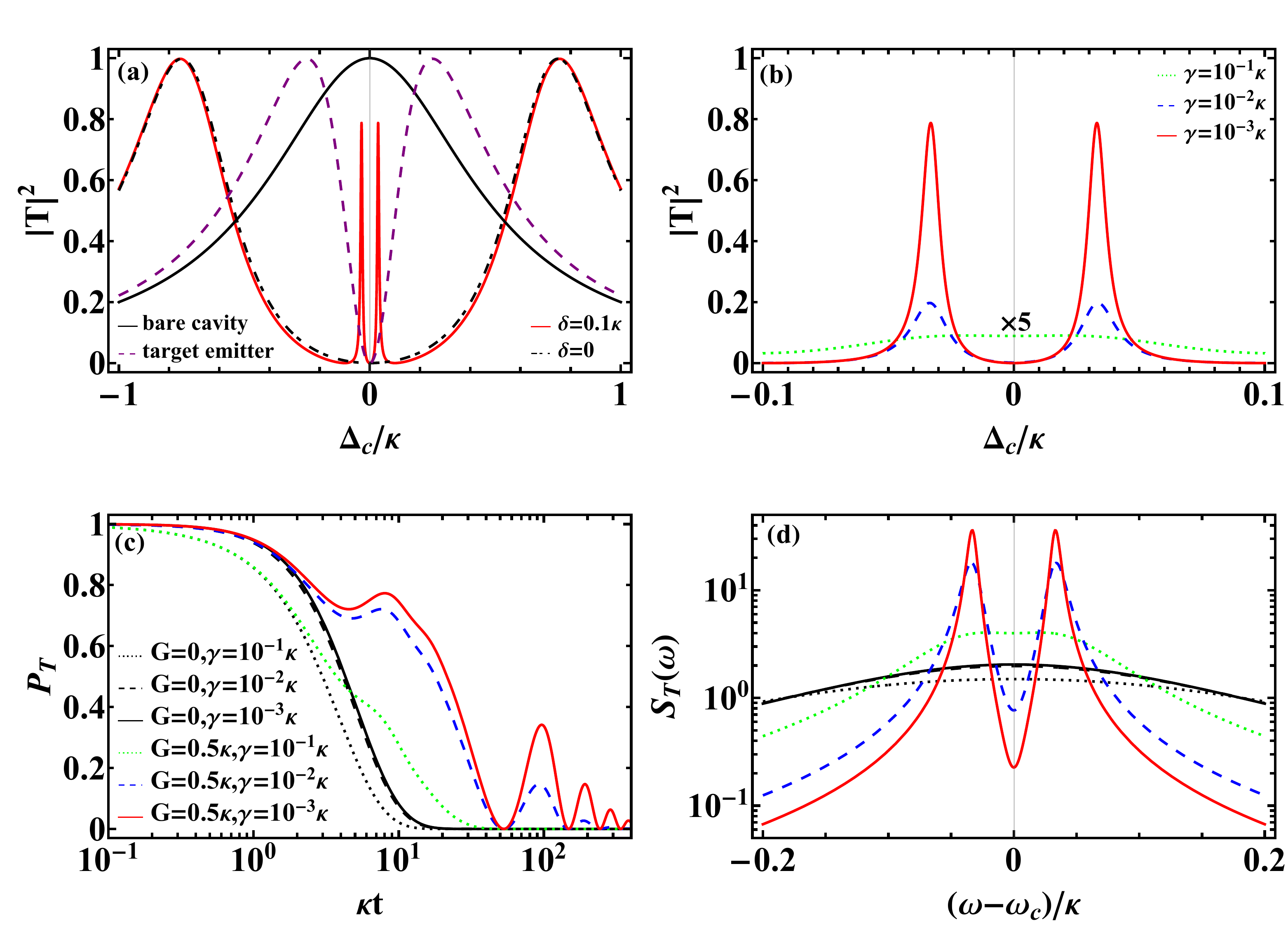}
 \caption{(a) Transmission spectrum $|T|^2$ of the coupled system with a target emitter with $G = 0.5\kappa$, $\gamma = 10^{-3}\kappa$ and $g = 0.25\kappa$. The red solid line corresponds to $\delta = 0.1\kappa$, and the black dashed line to $\delta = 0$. The black solid and purple dashed lines are the transmission spectra of the bare cavity and the cavity with a target emitter ($G = 0$), respectively. (b) Central transmission spectrum splitting under different values of $\gamma$ where $G = 0.5\kappa$ and $\delta = 0.1\kappa$. The population dynamics (c) and the spontaneous emission spectrum (d) of the target emitter under different values of $\gamma$ with $g=0.25\kappa$ and $\delta = 0.1\kappa$. Other parameters: $\gamma_T = \gamma$, $\delta_1 = -\delta_2 = \delta$ and $\delta_T = 0$. \label{Fig4}}
 \end{figure}

The total Hamiltonian including the target emitter is given by Eq. (\ref{system Hamiltonian}). The transmission of the coupled system is given by
\begin{equation}
T(\Delta_c)=\frac{\kappa /2}{i\Delta _{c}-\frac{\kappa }{2}+\sum_{j=1}^{3}%
\frac{\left \vert g_j\right \vert ^{2}}{i\left( \Delta _{c}-\delta _{j}\right)
-\frac{\gamma_{j} }{2}}},  \label{eq11transmission}
\end{equation}
where $g_1=g_2=G$, $\gamma_1=\gamma_2=\gamma$, $\gamma_{3}=\gamma_T$, and $\delta_3=\delta_{T}$. Assume that the target emitter is resonant with the cavity ($\delta_T=0$) with $\gamma_T = \gamma$ and $g = 0.25\kappa$. The transmission spectrum splits into two peaks, as shown by the purple dashed line in Fig. \hyperref[Fig4]{\ref{Fig4}(a)}. However, it is important to note that this spectral splitting does not necessarily indicate that the system has entered the strong coupling regime; it can also arise from quantum interference \cite{Liu2021PRB}. In contrast, introducing two non-identical auxiliary emitters into the cavity leads to two very sharp transmission peaks around the cavity frequency, which results from the suppression of the cavity linewidth (red solid line in Fig. \hyperref[Fig4]{\ref{Fig4}(a)}). When the two auxiliary emitters are identical, such sharp peaks disappear (black dashed line in Fig. \hyperref[Fig4]{\ref{Fig4}(a)}). As shown in Fig. \hyperref[Fig4]{\ref{Fig4}(b)}, the two splitting peaks become higher and sharper as $\gamma$ decreases, whereas they are barely resolved for $\gamma=0.1\kappa$ (greed dotted line, which has been amplified by 5 times for clarity).  Since the transmission spectrum alone is not a reliable indicator of strong coupling, we further calculated the population dynamics of the target emitter and the spontaneous emission spectrum of the system.

The dynamical evolution of the system can be obtained by solving the system
master equation  \cite{Scully1997-book,Wei2025}
\begin{align}
\dot{\rho}=& i\left[ \rho ,\hat{H}\right]+\sum_{i=1}^{3}\frac{\gamma_i }{2}\left( 2\sigma _{i}\rho \sigma _{i}^{\dagger
}-\sigma _{i}^{\dagger }\sigma _{i}\rho -\rho \sigma _{i}^{\dagger }\sigma
_{i}\right) \nonumber \\ &+\frac{\kappa}{2}\left( 2a\rho a^{\dagger
}-a^{\dagger }a\rho -\rho a^{\dagger }a\right),
\label{master-equation}
\end{align}
where $H$ is given by Eq. (\ref{system Hamiltonian}), $\hat{\sigma}_3=\hat{\sigma}_{T}$ and $\gamma_{3}=\gamma_{T}$.
The excitation probability of the target emitter $P_T(t)=Tr[\sigma^{\dagger}_{T}\sigma_{T}\rho(t)]$. 

In the numerical simulation, the target emitter is assumed to be resonant with the cavity frequency. Initially, only the target emitter is excited, while the auxiliary emitters are in their ground states and the cavity is in the vacuum state. 
Figure \hyperref[Fig4]{\ref{Fig4}(c)} shows $P_{T}(t)$ for three different values of $\gamma$: colored lines represent the results with auxiliary emitters ($G\neq0$), and black lines correspond to the case without auxiliary emitters($G=0$).
The results strikingly highlight the advantages of our scheme. When no auxiliary emitters are present ($G = 0$), $P_{T}(t)$ decays exponentially without oscillations for all three values of $\gamma$ (black lines in Fig. \hyperref[Fig4]{\ref{Fig4}(c)}). The emitter lifetimes remain approximately the same even though the values of $\gamma$ vary by orders of magnitude. This is because, in the weak-coupling regime of the cavity-emitter system, the effective decay rate of the emitter is approximately $\Gamma_{eff}\approx 2g^2/\kappa$, which is independent of $\gamma$ \cite{PhysRevResearch.6.L012050}. In contrast, when auxiliary emitters are introduced ($G \neq 0$), clear Rabi oscillations emerge--a hallmark of strong coupling--and the target emitter lifetime is significantly prolonged [blue dashed line ($\gamma=0.01\kappa $) and red solid line ($\gamma=0.001\kappa$) and more results are shown in the supplementary material]. The decrease of $\gamma$ enhances both the excitation lifetime and the oscillation amplitude, consistent with our earlier observation that a smaller $\gamma$ reduces the effective cavity linewidth (Fig. \hyperref[Fig2]{\ref{Fig2}(b)}). The initial rapid drop in excitation arises from fast energy loss via the cavity decay channel. Beyond this transient regime, the excitation energy becomes predominantly trapped within the target-emitter–auxiliary-emitter subsystem. 

The spontaneous emission spectrum is an alternative physical quantity to verify the occurrence
of strong coupling. When strong coupling arises between the
emitter and the system, the spontaneous emission spectrum exhibits
splitting. The normalized spontaneous emission spectrum is given by \cite{57-Pelton_Storm_Leng_2019,58-Huang_Wang_Liang_Yu_2021}
\begin{eqnarray}
S_T\left( \omega \right)  &=&\frac{\int dt_{1}\int
dt_{2}e^{i\omega \left( t_{2}-t_{1}\right) }\left \langle \sigma _{T}^{\dagger }\left(
t_{1}\right) \sigma _{T}\left( t_{2}\right) \right \rangle}{2\pi
\int dt\left \langle \sigma _{T}^{\dagger }\left( t\right) \sigma _{T}\left(
t\right) \right \rangle} \label{spontaneous emission spectrum}
\end{eqnarray}%
where the two-time correlation function $\left \langle \sigma _{T}^{\dagger }\left(
t_{1}\right) \sigma _{T}\left( t_{2}\right) \right \rangle$ can be calculated by the quantum regression theorem \cite{Scully1997-book,Liao2012}. The detail expression for Eq. (\ref{spontaneous emission spectrum}) is given in the Supplementary material. 



For a weakly coupled cavity QED system where $g<(\kappa+\gamma)/4$, the spontaneous emission spectrum of the target emitter exhibits a single-peak structure (as shown by the black lines in Fig. \hyperref[Fig4]{\ref{Fig4}(d)} for three different values of $\gamma$). By introducing two non-identical auxiliary emitters, the spontaneous emission spectrum splits into two peaks for all three values of $\gamma$. When $\gamma=0.1\kappa$, the spectrum splitting is barely resolved (green dotted line). In contrast,  smaller $\gamma$ leads to  narrower spectral linewidth and  higher splitting peak, which corresponds to improved oscillation amplitude and increased lifetimes in the time domain (see the blue dashed and red solid lines in Fig. \hyperref[Fig4]{\ref{Fig4}(d)}). 

\prlsection{Conclusion} We propose a hybrid system consisting of a bad cavity and two non-identical auxiliary emitters to reduce the effective cavity linewidth, thereby converting a weakly coupled cavity QED system into an effective strongly coupled one. The auxiliary emitters, symmetrically detuned from the cavity, create a subradiant polariton mode with a linewidth significantly narrower than that of the bare cavity. Tuning the system parameters enhances the dark-state component, resulting in sharp transmission peaks. Placing a target emitter into this hybrid cavity yields two sharp peaks in the transmission spectrum. Furthermore, the target emitter exhibits prolonged Rabi oscillations in its population dynamics, and its spontaneous emission spectrum shows clear splitting—both confirming the achievement of strong coupling between the target emitter and the hybrid cavity. Our results may find important applications in quantum computation and quantum sensing.

\prlsection{\label{sec:acknowledgments}Acknowledgments}
This work was supported by the National Key R\&D Program of China (Grant No. 2021YFA1400800), the Key Program of National Natural Science Foundation of China (Grant No. 12334017), Guangdong Provincial Quantum Science Strategic Initiative (Grant No. GDZX2505001 and GDZX2406001), and Guangdong Basic and Applied Basic Research Foundation (Grant No. 2026A1515011705).

\noindent \textbf{Disclosures.} The authors declare that there are no
conflicts of interest related to this article.\newline

\noindent \textbf{Data availability.} Data underlying the results presented
in this paper may be obtained from the authors upon reasonable request.




%


\clearpage
\widetext

\begin{center}
\textbf{Supplementary material for ``Beating the Bad-Cavity Limit via Auxiliary-Emitter Linewidth Squeezing''}
\end{center}

\setcounter{equation}{0}
\setcounter{figure}{0}
\setcounter{page}{1}
\makeatletter
\renewcommand{\theequation}{S\arabic{equation}}
\renewcommand{\thefigure}{S\arabic{figure}}
\renewcommand{\bibnumfmt}[1]{[S#1]}

\section{Derivation of the cavity transmission spectrum}

Here, we present a general and universal expression for the transmission
spectrum of a system consisting of $n$ two-level emitters coupled to a
single-mode cavity field. The free part of system Hamiltonian reads ($\hbar
=1$)%
\begin{equation}
H_{0}=\sum_{j=1}^{n}\omega _{j}\sigma _{j}^{\dag}\sigma _{j}+\omega
_{c}a_{c}^{\dag }a_{c},
\end{equation}%
and the interaction part reads%
\begin{equation}
H_{int}=\sum_{j=1}^{n}g_{j}\sigma _{j}^{\dag}a_{c}+H.c.
\end{equation}%
Where $\omega _{j}$ and $\omega _{c}$ are the transition frequency of
the $v$-th emitter and the cavity field frequency, respectively, and $g_{j}$
is the coupling strength between the $v$-th emitter and the cavity field.
Considering the leakage from both ends of the cavity mirrors and the
dissipation of the emitters into free space, the total Hamiltonian of the
system, including the external environment $b_{\omega }$, is%
\begin{eqnarray}
H &=&H_{0}+H_{int}+\int_{-\infty }^{+\infty }d\omega \omega b_{\omega
}^{\dag }b_{\omega }+\left( \int_{-\infty }^{+\infty }d\omega \left( \kappa
_{L\omega }+\kappa _{R\omega }\right) a_{c}b_{\omega }^{\dag }+H.c.\right)
\nonumber \\
&&+\sum_{j=1}^{n}\left( \int_{-\infty }^{+\infty }d\omega g_{v\omega }\sigma
_{j}b_{\omega }^{\dag }+H.c.\right) 
\end{eqnarray}%
where $\kappa _{L\omega }$ ($\kappa _{R\omega }$) and $g_{v\omega }$ are the
coupling strengths of the left (right) end of the cavity mirror and the $v$%
-th emitter to the reservoir, respectively. Hence, quantum Langevin equation
of system reads
\begin{equation}
\frac{d}{dt}O\left( t\right) =\frac{i}{\hbar }[H,O\left( t\right) ].
\end{equation}%
Under weak-excitation $\sigma _{j}^{z}\left( t\right) \simeq -1,$ we obtain%
\begin{eqnarray}
\frac{d}{dt}\sigma _{j}\left( t\right)  &\simeq &-i\omega _{j}\sigma
_{j}\left( t\right) -ig_{j}a_{c}\left( t\right) -\frac{\gamma _{j}}{2}\sigma
_{j}\left( t\right) ,  \nonumber \\
\frac{d}{dt}a_{c}\left( t\right)  &=&-i\omega _{c}a_{c}\left( t\right)
-i\sum_{j=1}^{n}g_{j}^{\ast }\sigma _{j}\left( t\right) -\frac{\kappa _{1}+\kappa
_{2}}{2}a_{c}\left( t\right) -i\sqrt{\kappa _{1}}L_{in}\left( t\right) +i%
\sqrt{\kappa _{2}}R_{in}\left( t\right) ,  \nonumber \\
\frac{d}{dt}a_{c}\left( t\right)  &=&-i\omega _{c}a_{c}\left( t\right)
-i\sum_{j=1}^{n}g_{j}^{\ast }\sigma _{j}\left( t\right) +\frac{\kappa _{1}-\kappa
_{2}}{2}a_{c}\left( t\right) +i\sqrt{\kappa _{1}}L_{out}\left( t\right) +i%
\sqrt{\kappa _{2}}R_{in}\left( t\right) ,  \nonumber \\
\frac{d}{dt}a_{c}\left( t\right)  &=&-i\omega _{c}a_{c}\left( t\right)
-i\sum_{j=1}^{n}g_{j}^{\ast }\sigma _{j}\left( t\right) +\frac{\kappa _{1}+\kappa
_{2}}{2}a_{c}\left( t\right) +i\sqrt{\kappa _{1}}L_{out}\left( t\right) -i%
\sqrt{\kappa _{2}}R_{out}\left( t\right) .  \label{S5}
\end{eqnarray}%
Here, $\kappa _{1}$ and $\kappa _{2}$\ are the leakage rates of the left and
right ends of the cavity mirror, respectively. The inputs (outputs) at the
two ends of the cavity mirror are $L_{in}\left( t\right) $ ($L_{out}\left(
t\right) $) and $R_{in}\left( t\right) $ ($R_{out}\left( t\right) $),
respectively, and the loss of the $v$-th emitter into the reservoir is $%
\gamma _{j}.$ Based Eqs. (\ref{S5}), employing Fourier transformation $O\left(
t\right) =\frac{1}{\sqrt{2\pi }}\int_{-\infty }^{+\infty }e^{-i\omega
t}O\left( \omega \right) d\omega ,$ we have
\begin{eqnarray}
\sigma _{j}\left( \omega \right)  &=&\frac{ig_{j}}{i\left( \omega
-\omega _{j}\right) -\frac{\gamma _{j}}{2}}a_{c}\left( \omega \right) ,
\nonumber \\
a_{c}\left( \omega \right)  &=&\frac{i\sqrt{\kappa _{1}}L_{in}\left( \omega
\right) -i\sqrt{\kappa _{2}}R_{in}\left( \omega \right) }{i\left( \omega
-\omega _{c}\right) -\frac{\kappa _{1}+\kappa _{2}}{2}+\sum_{j=1}^{n}\frac{%
\left\vert g_{j}\right\vert ^{2}}{i\left( \omega -\omega _{j}\right) -%
\frac{\gamma _{j}}{2}}},  \nonumber \\
L_{out}\left( \omega \right)  &=&i\sqrt{\kappa _{1}}a_{c}\left( \omega
\right) -L_{in}\left( \omega \right) ,  \nonumber \\
R_{out}\left( \omega \right)  &=&-i\sqrt{\kappa _{2}}a_{c}\left( \omega
\right) -R_{in}\left( \omega \right) .  \label{S6}
\end{eqnarray}%
Considering that a weak probe field is input from the left end of the cavity
mirror while there is no input at the right end, i.e., $R_{in}\left( \omega
\right) =0,$ the transmission coefficient of the system is%
\begin{eqnarray}
T &=&\frac{R_{out}\left( \omega \right) }{L_{in}\left( \omega \right) }
\nonumber \\
&=&\frac{\sqrt{\kappa _{1}\kappa _{2}}}{i\Delta _{c}-\frac{\kappa
_{1}+\kappa _{2}}{2}+\sum_{j=1}^{n}\frac{\left\vert g_{j}\right\vert ^{2}}{i\left(
\Delta _{c}-\delta _{j}\right) -\frac{\gamma _{j}}{2}}},  \label{S7}
\end{eqnarray}%
where $\Delta _{c}=\omega -\omega _{c}$ and $\delta _{j}=\omega
_{j}-\omega _{c}$ are the detunings of the probe field and the $v$-th
emitters from the cavity field, respectively. Employing Eq. (\ref{S7}), we can obtain Eqs. (6) and (10), respectively.

\begin{figure}[tbh]
\centering
\includegraphics[width=0.95\columnwidth]{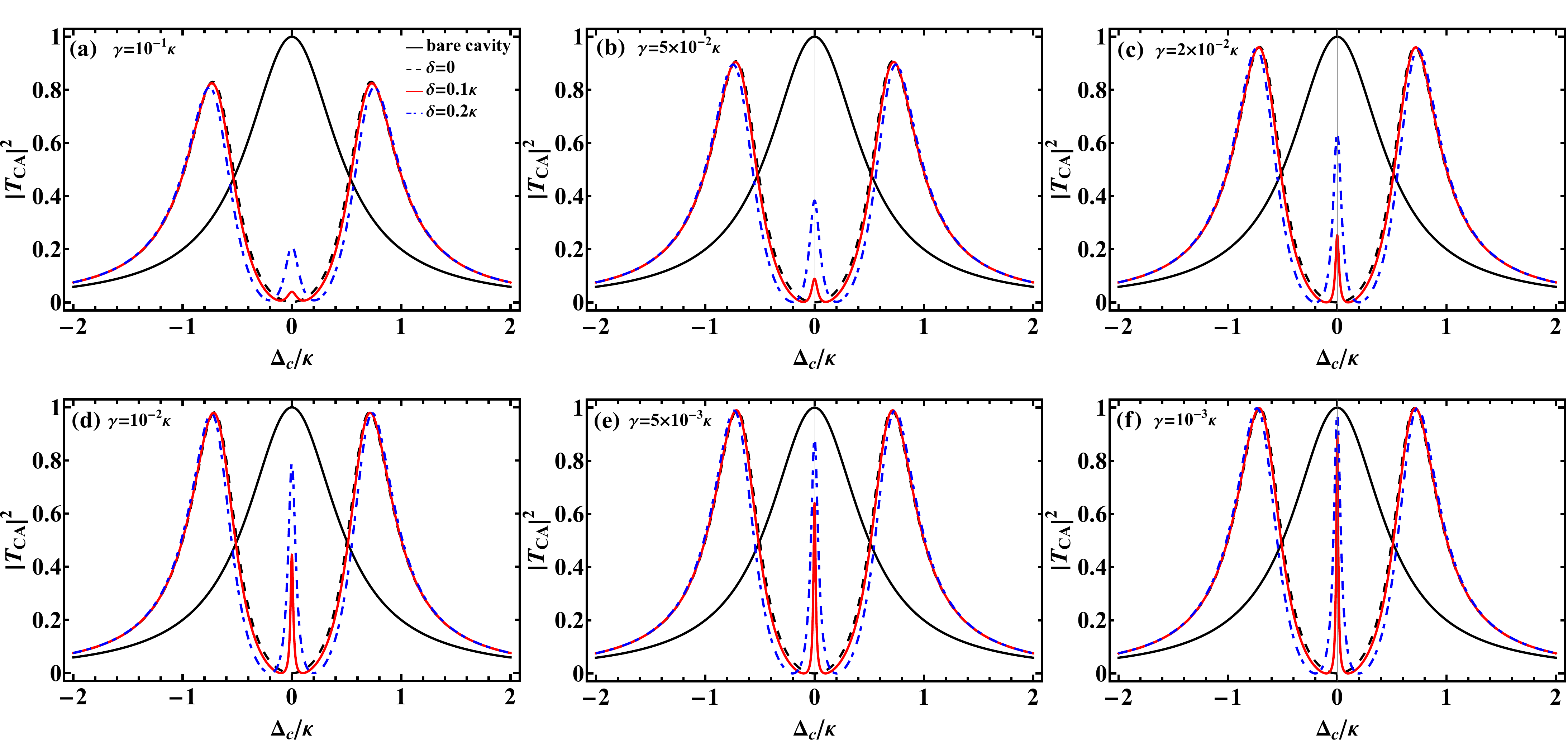}
\caption{Transmission spectra for a bad cavity with two non-identical auxiliary emitters for different values of $\gamma$ and $\delta$. (a) $\gamma=0.1\kappa$; (b) $\gamma=0.05\kappa$; (c) $\gamma=0.02\kappa$; (d) $\gamma=0.01\kappa$; (e) $\gamma=0.005\kappa$; (f) $\gamma=0.001\kappa$. For comparison, the transmission spectrum of bare cavity is also shown (black solid line). Other parameters: $G=0.5\protect \kappa$ and $\delta_1 = -\delta_2 = \delta.$} \label{FS1}
\end{figure}

In Figs. \ref{FS1}, we present the transmission spectra of the hybrid cavity system for different values of $\gamma$, where the detunings of the two auxiliary emitters from the cavity field frequency satisfy $\delta_1 = \delta_2 = \delta$. By adjusting $\delta$ and $\gamma$, both the linewidth and the peak height of the central peak can be modified. In other words, when $\gamma$ is relatively large, an appropriate increase in $\delta$ allows the central peak to maintain a certain height while exhibiting a narrow linewidth; when $\gamma$ is small, a suitable decrease in $\delta$ has little effect on the peak height but can further reduce the linewidth.

\section{Analytical approximation of the cavity transmission spectrum}

Here, we derive an approximate expression for the transmission spectrum of the hybrid cavity around the resonant frequency. Let 
\begin{equation}
f(\Delta_c)=\sum_{j=1}^{2}\frac{\left \vert G\right \vert ^{2}}{i\left( \Delta _{c}-\delta _{j}\right)-\frac{\gamma }{2}}.
\end{equation}
When $\delta_{1}=-\delta_{2}=\delta$, we have
\begin{equation}
f(\Delta_c)=\frac{|G|^2(2i\Delta_c-\gamma)}{(i\Delta_{c}-\frac{\Gamma}{2})+\delta^2}.
\end{equation}
Then, by expanding $f(\Delta_c)$ around $\Delta_c\approx 0$, we can obtain
\begin{equation}
f(\Delta_{c})\approx -\frac{|G|^2\gamma}{\delta^2+\frac{\gamma^2}{4}}+2i|G|^2\frac{\delta^2-\frac{\gamma^2}{4}}{(\delta^2+\frac{\gamma^2}{4})^2}\Delta_c+\mathcal{O}(\Delta_c^2)
\end{equation}
It is then readily obtain from Eq. (7) that
\begin{equation}
T_{ca}=\frac{\kappa\eta/2}{i\Delta_{c}-\frac{\Gamma_{eff}}{2}}
\end{equation}
where 
\begin{equation}
\Gamma_{eff}=\Big(\kappa+\frac{2|G|^2\gamma}{\delta^2+\frac{\gamma^2}{4}}\Big)\eta
\end{equation}
is the effective linewidth of the hybrid cavity and 
\begin{equation}
\eta=\frac{(\delta^2+\frac{\gamma^2}{4})^2}{(\delta^2+\frac{\gamma^2}{4})^2+2|G|^2(\delta^2-\frac{\gamma^2}{4})}.
\end{equation}

\section{Eigenvalue and eigenstate of the effective Hamiltonian}

In the single-excitation subspace, the eigenstate $\left\vert \Psi
_{CA}\right\rangle =$ $\theta _{1}\left\vert e,g,0\right\rangle +\theta
_{2}\left\vert g,e,0\right\rangle +\theta _{3}\left\vert g,g,1\right\rangle $
of the hybrid cavity can be derived by solve the the energy eigenvalue
equation $H_{CA}^{eff}\left\vert \Psi _{CA}\right\rangle =E\left\vert \Psi
_{CA}\right\rangle .$ From the above discussion, it can be seen that for the
case where $\delta _{1}=-\delta _{2}=\delta ,$ the transmission spectrum has
an extremum at $\Delta _{c}=0;$ that is, the energy $E$ corresponding to the
central peak is purely imaginary, i.e., $E=-i\frac{\Delta _{fw}}{2}$. Here, $%
\Delta _{fw}$ is the linewidth of the eigenstate $\left\vert \Psi
_{CA}^{c}\right\rangle $ corresponding to the central peak. By solving the
eigenvalue equation $H_{CA}^{eff}\left\vert \Psi _{CA}\right\rangle
=E\left\vert \Psi _{CA}\right\rangle ,$ we have
\begin{equation}
\det \left(
\begin{array}{ccc}
\delta -\frac{i\gamma }{2}-E & 0 & G \\
0 & -\delta -\frac{i\gamma }{2}-E & G \\
G^{\ast } & G^{\ast } & -\frac{i\kappa }{2}-E%
\end{array}%
\right) =0. 
\end{equation}%
Hence, the energies of the hybrid cavity can be derived as
\begin{eqnarray}
E_{1} &=&-\frac{1}{3}\left( x+\frac{Q}{H}-H\right) , \\
E_{2,3} &=&-\frac{1}{3}\left( x-\frac{Q}{H}e^{\pm \frac{i\pi }{3}}+He^{\mp
\frac{i\pi }{3}}\right) , 
\end{eqnarray}%
with%
\begin{eqnarray}
x &=&i\left( \gamma +\frac{\kappa }{2}\right) , \\
y &=&-\left( 2G^{2}+\frac{\gamma ^{2}}{4}+\delta ^{2}+\frac{\gamma \kappa }{2%
}\right) , \\
z &=&-i\left( G^{2}\gamma +\frac{\gamma ^{2}\kappa }{8}+\frac{\delta
^{2}\kappa }{2}\right) , \\
Q &=&3y-x^{2}, \\
W &=&2x^{3}-9xy+27z, \\
R &=&3\left( 2zW+y^{2}Q+y^{3}-27z^{2}\right) , \\
H &=&\left( \frac{3}{2}\sqrt{R}-\frac{W}{2}\right) ^{\frac{1}{3}}. 
\end{eqnarray}%
As shown in Figs. 3(a-d), the energies of the hybrid cavity are $E_1 = E_+,$ $E_2 = E_-,$ and $E_3 = E_0,$ respectively. Where the energy of the eigenstate $\left\vert \Psi
_{CA}^{c}\right\rangle$ corresponding to the central peak is $E_{0},$ with a vanished real part. 

Furthermore, we can obtain the eigenstate $\left\vert \Psi
_{CA}^{c}\right\rangle $ by solving the equation
\begin{equation}
\left(
\begin{array}{ccc}
\delta -\frac{i\gamma }{2} & 0 & G \\
0 & -\delta -\frac{i\gamma }{2} & G \\
G^{\ast } & G^{\ast } & -\frac{i\kappa }{2}%
\end{array}%
\right) \left(
\begin{array}{c}
\theta _{1} \\
\theta _{2} \\
\theta _{3}%
\end{array}%
\right) =-i\frac{\Delta _{fw}}{2}\left(
\begin{array}{c}
\theta _{1} \\
\theta _{2} \\
\theta _{3}%
\end{array}%
\right) .  \label{S24}
\end{equation}%
Here, $\Delta _{fw}=-2\text{Im}\left( E_{0}\right) $ is the linewidth of the
eigenstate $\left\vert \Psi _{CA}^{c}\right\rangle .$ Based on Eq. (\ref{S24}) and
the normalization condition, we have%
\begin{eqnarray}
\left( \delta +i\frac{\Delta _{fw}-\gamma }{2}\right) \theta _{1}+G\theta
_{3} &=&0,  \nonumber \\
\left( -\delta +i\frac{\Delta _{fw}-\gamma }{2}\right) \theta _{2}+G\theta
_{3} &=&0,  \nonumber \\
G^{\ast }\left( \theta _{1}+\theta _{2}\right) +i\frac{\Delta _{fw}-\kappa }{%
2}\theta _{3} &=&0,  \nonumber \\
\left\vert \theta _{1}\right\vert ^{2}+\left\vert \theta _{2}\right\vert
^{2}+\left\vert \theta _{3}\right\vert ^{2} &=&1.  \label{x2}
\end{eqnarray}%
It is noted that when $\delta =0$, the eigenstate $\left\vert \Psi
_{CA}^{c}\right\rangle $ can be drived as a dark state $\left\vert \Psi
_{CA}^{0}\right\rangle =\left( \left\vert e,g,0\right\rangle -\left\vert
g,e,0\right\rangle \right) /\sqrt{2}\,$with the linewidth $\Delta
_{fw}=\gamma .$ In general, the eigenstate reads%
\begin{equation}
\left\vert \Psi _{CA}^{c}\right\rangle =\frac{-4G\delta +2iG\left( \Delta
_{fw}-\gamma \right) }{4\delta ^{2}+\left( \Delta _{fw}-\gamma \right) ^{2}}%
\theta _{3}\left\vert e,g,0\right\rangle +\frac{4G\delta +2iG\left( \Delta
_{fw}-\gamma \right) }{4\delta ^{2}+\left( \Delta _{fw}-\gamma \right) ^{2}}%
\theta _{3}\left\vert g,e,0\right\rangle +\theta _{3}\left\vert
g,g,1\right\rangle ,  \label{x3}
\end{equation}%
with%
\begin{equation}
\theta _{3}=\left( \frac{8\left\vert G\right\vert ^{2}}{4\delta ^{2}+\left(
\Delta _{fw}-\gamma \right) ^{2}}+1\right) ^{-\frac{1}{2}}.  \label{x4}
\end{equation}%
By combining the real and imaginary parts of the probability amplitudes
corresponding to the basis vectors of the eigenstate $\left\vert \Psi
_{CA}^{c}\right\rangle $, we can obtain%
\begin{equation}
\left\vert \Psi _{CA}^{c}\right\rangle =\frac{\alpha _{D}}{\sqrt{2}}\left(
\left\vert e,g,0\right\rangle -\left\vert g,e,0\right\rangle \right) +\frac{%
\alpha _{B_{1}}}{\sqrt{2}}\left( \left\vert e,g,0\right\rangle +\left\vert
g,e,0\right\rangle \right) +\alpha _{B_{2}}\left\vert g,g,1\right\rangle ,
\label{x5}
\end{equation}%
where the probability amplitude of the dark state component reads%
\begin{equation}
\alpha _{D}=\frac{-4\sqrt{2}G\delta }{4\delta ^{2}+\left( \Delta
_{fw}-\gamma \right) ^{2}}\theta _{3},  \label{x6}
\end{equation}%
and the probability amplitude of the bright state components read%
\begin{eqnarray}
\alpha _{B_{1}} &=&\frac{2\sqrt{2}iG\left( \Delta _{fw}-\gamma \right) }{%
4\delta ^{2}+\left( \Delta _{fw}-\gamma \right) ^{2}}\theta _{3},  \nonumber
\\
\alpha _{B_{2}} &=&\theta _{3}.  \label{x7}
\end{eqnarray}%
Here, the linewidth $\Delta _{fw}=-2\text{Im}\left(E_{0}\right) \approx
\kappa _{eff},$ as shown in Figs. 3(c-d). Figs. 3(e-f) depict the probabilities of dark and bright components which
are defined as $\left\vert \alpha _{D}\right\vert ^{2}$ and $\left\vert
\alpha _{B}\right\vert ^{2}=\left\vert \alpha _{B_{1}}\right\vert
^{2}+\left\vert \alpha _{B_{2}}\right\vert ^{2},$ respectively. Parameters $G,$ and $\delta$ can alter the weights of the bright and the dark
components. It is noted that for the case of $\delta
=0,$ the eigenstate $\left\vert \Psi _{CA}^{c}\right\rangle $ reduce to the
dark state $\left\vert \Psi _{CA}^{0}\right\rangle =\frac{\sqrt{2}}{2}\left(
\left\vert e,g,0\right\rangle -\left\vert g,e,0\right\rangle \right) $ which
corresponding to the central peak disappears, the linewidth reduces to $%
\Delta _{fw}=\gamma .$

\section{System dynamics evolution and Spontaneous emission spectrum}

Calculating the dynamical evolution of the system is an effective method to
validate the strength of coupling between emitters and optical systems. For
example, when an emitter couples to an optical cavity, the evolution of the
emitter population exhibits rapid exponential decay in the weak-coupling
regime, while Rabi oscillations occur in the strong-coupling regime.
Generally speaking, under actual conditions, it is difficult to control the
cavity loss at a low level, which will have a huge negative effect on the
light-matter interaction effect of the cavity qed system, such as
significantly reducing the duration of the interaction. In this section, our
objective is to utilize two auxiliary emitters-cavity hybrid system shown in
Fig. 1(a) to achieve a strong coupling effect between the target emitter and
the optical system under the condition of a bad cavity ($g<\left( \kappa
+\gamma \right) /4$).
\begin{figure}[tbh]
\centering
\includegraphics[width=0.95\columnwidth]{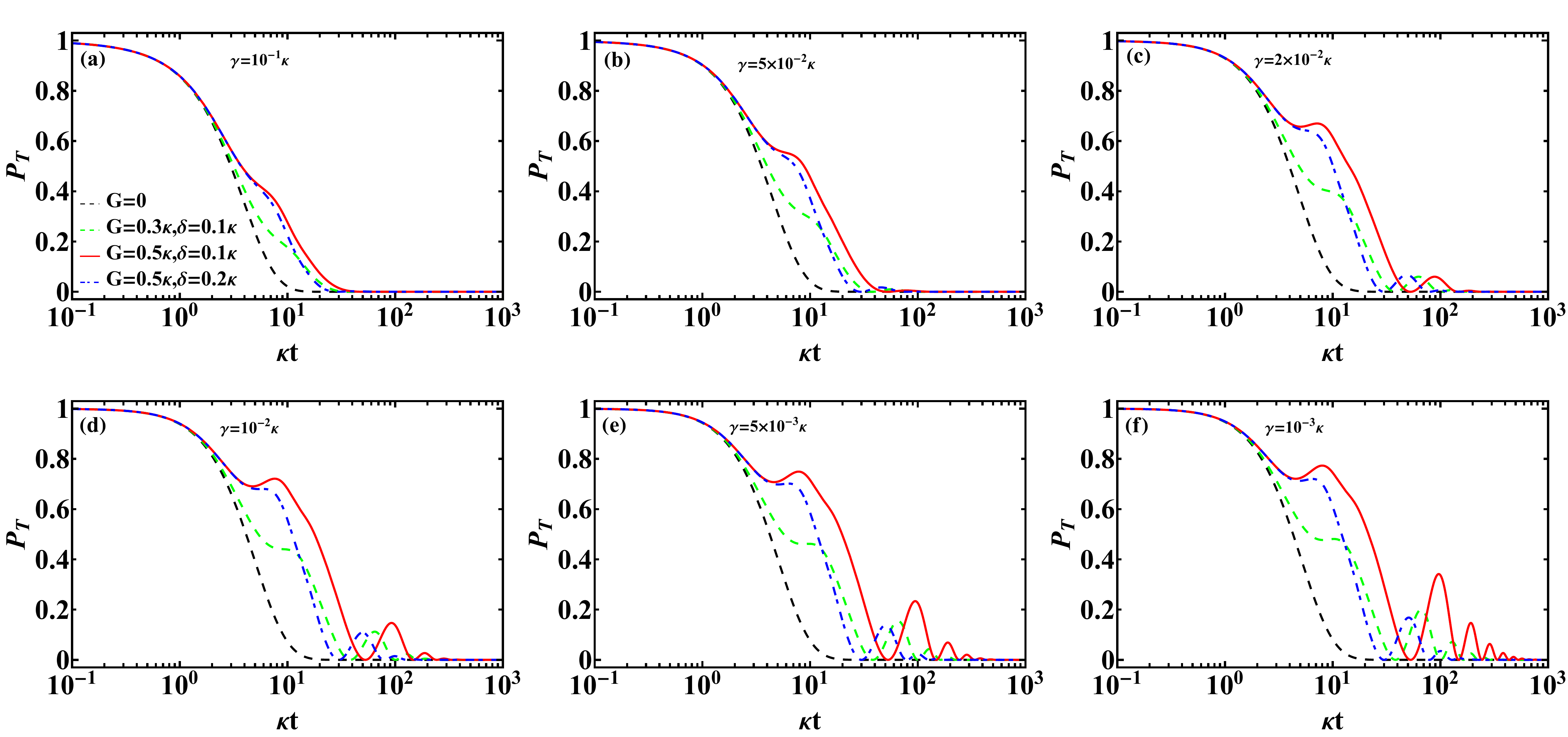}
\caption{The population dynamics of the target emitter for different values of under different values of $\gamma$, $G$ and $\delta$.  (a) $\gamma=0.1\kappa$; (b) $\gamma=0.05\kappa$; (c) $\gamma=0.02\kappa$; (d) $\gamma=0.01\kappa$; (e) $\gamma=0.005\kappa$; (f) $\gamma=0.001\kappa$.  Other parameters: $g = 0.25\kappa$ and $\delta_T = 0$.} \label{FS2}
\end{figure}

For the sake of discussion, the loss of all emitters into the reservoir is $\gamma_{1,2,3} = \gamma$. The dynamical evolution $\left \langle \sigma _{j}^{\dag }\sigma
_{j}\right \rangle $ of the system can be obtained by solving the system
master equation $\dot{\rho}=\frac{i}{\hbar }\left[ \rho ,H\right]
+\sum_{v=1}^{3}\frac{\gamma }{2}\left( 2\sigma _{j}\rho \sigma _{j}^{\dag
}-\sigma _{j}^{\dag }\sigma _{j}\rho -\rho \sigma _{j}^{\dag }\sigma
_{j}\right) +\frac{\kappa }{2}\left( 2a\rho a^{\dag }-a^{\dag }a\rho -\rho
a^{\dag }a\right) $. Here, for the sake of convenience in writing, we denote $\left \langle \sigma _{T}^{\dag }\sigma
_{T}\right \rangle = \left \langle \sigma _{3}^{\dag }\sigma
_{3}\right \rangle $. However, for multi-atom open cavity QED systems,
pursuing analytical solutions is a difficult task. Here, we introduce
another method for calculating the dynamic evolution of the system: under
the low-excitation approximation $\sigma _{z}\simeq -1$, the
Heisenberg-Langevin equations of the system are identical to the equations
describing the classical field amplitudes \cite{56-PhysRevA.77.063833}. In our model, the
Heisenberg-Langevin equations read%
\begin{eqnarray}
\frac{d}{dt}\sigma _{j}\left( t\right)  &=&\left( -i\omega _{j}-\frac{%
\gamma }{2}\right) \sigma _{j}\left( t\right) -i\chi _{j}a_{c}\left(
t\right) ,  \label{15} \\
\frac{d}{dt}a_{c}\left( t\right)  &=&\left( -i\omega _{c}-\frac{\kappa }{2}%
\right) a_{c}\left( t\right) -i\sum_{v=1v}^{3}\chi _{j}^{\ast }\sigma
_{j}\left( t\right) ,  \label{16}
\end{eqnarray}%
where $\chi _{1,2}=G$ and $\chi _{3}=g$ are the coupling coefficients of the
auxiliary emitters and the target emitter with the cavity field,
respectively. Under the low excitation approximation $\sigma _{z}\simeq -1,$
the system described by Eqs. (\ref{15},\ref{16}) is identical to the system
described by three independent cavities $C_{j}$ coupled to a common cavity $A
$. Therefore, Eqs. (\ref{15},\ref{16}) can be considered as the evolution
equation for the classical field amplitudes, i.e.,
\begin{equation}
\frac{d}{dt}\left(
\begin{array}{c}
\tilde{C}_{1} \\
\tilde{C}_{2} \\
\tilde{C}_{3} \\
\tilde{A}%
\end{array}%
\right) =M\left(
\begin{array}{c}
\tilde{C}_{1} \\
\tilde{C}_{2} \\
\tilde{C}_{3} \\
\tilde{A}%
\end{array}%
\right) ,  \label{17}
\end{equation}%
where the field amplitudes are defined as $\tilde{O}\left( t\right) =O\left(
t\right) e^{i\omega _{c}t},$ and
\begin{equation}
M=\left(
\begin{array}{cccc}
\tilde{\omega}_{1} & 0 & 0 & -iG \\
0 & \tilde{\omega}_{2} & 0 & -iG \\
0 & 0 & \tilde{\omega}_{3} & -ig \\
-iG^{\ast } & -iG^{\ast } & -ig^{\ast } & -\frac{\kappa }{2}%
\end{array}%
\right) .  \label{18}
\end{equation}%
Here, the complex frequency $\tilde{\omega}_{j}$ are defined as $\tilde{%
\omega}_{j}=-i\delta _{j}-\frac{\gamma }{2}.$ The solutions of $\tilde{O}%
\left( t\right) =\sum_{j=1}^{4}B_{j}e^{-i\lambda _{j}t}$ are linear
combinations of exponential functions at the eigenfrequencies $\lambda _{j},$
where $\lambda _{j}$ are the roots of the equation $\det \left( M+i\lambda
\right) =0.$ Taking into account the initial conditions $\tilde{C}_{3}\left(
0\right) =1,$ and $\tilde{C}_{1,2}\left( 0\right) =\tilde{A}\left( 0\right)
=0,$ we have
\begin{eqnarray}
\lambda _{1,2} &=&-\frac{b}{4}-\varpi \pm \frac{1}{2}\sqrt{2p-4\varpi ^{2}+%
\frac{q}{\varpi }},  \notag \\
\lambda _{3,4} &=&-\frac{b}{4}+\varpi \pm \frac{1}{2}\sqrt{2p-4\varpi ^{2}-%
\frac{q}{\varpi }}.  \label{19}
\end{eqnarray}%
Where the parameters in $\lambda _{j}$ are defined as%
\begin{eqnarray}
p &=&\frac{3b^{2}-8c}{8},  \notag \\
q &=&\frac{b^{3}-4bc+8d}{8},  \notag \\
r &=&c^{2}-3bd+12e,  \notag \\
s &=&2c^{3}-9c\left( bd+8e\right) +27\left( d^{2}+b^{2}e\right) ,  \notag \\
\xi  &=&\sqrt[3]{\frac{s+\sqrt{s^{2}-4r^{3}}}{2}},  \notag \\
\varpi  &=&\frac{1}{2}\sqrt{\frac{1}{3}\left( 2p+\xi +\frac{r}{\xi }\right) }%
.  \label{20}
\end{eqnarray}%
with%
\begin{eqnarray}
b &=&i\left( \frac{\kappa }{2}-\sum_{j}\tilde{\omega}_{j}\right) ,  \notag \\
c &=&\frac{\sum_{j}\kappa \tilde{\omega}_{j}}{2}-\left( 2\left \vert
G\right \vert ^{2}+\left \vert g\right \vert ^{2}\right) -\tilde{\omega}_{1}%
\tilde{\omega}_{2}-\left( \tilde{\omega}_{1}+\tilde{\omega}_{2}\right)
\tilde{\omega}_{3},  \notag \\
d &=&i[\left \vert g\right \vert ^{2}\left( \tilde{\omega}_{1}+\tilde{\omega}%
_{2}\right) +\left \vert G\right \vert ^{2}\left( \tilde{\omega}_{1}+\tilde{%
\omega}_{2}+2\tilde{\omega}_{3}\right)   \notag \\
&&+\tilde{\omega}_{1}\tilde{\omega}_{2}\tilde{\omega}_{3}-\frac{\kappa }{2}%
\left( \tilde{\omega}_{1}\tilde{\omega}_{2}+\tilde{\omega}_{1}\tilde{\omega}%
_{3}+\tilde{\omega}_{2}\tilde{\omega}_{3}\right) ],  \notag \\
e &=&\left \vert g\right \vert ^{2}\tilde{\omega}_{1}\tilde{\omega}%
_{2}+\left \vert G\right \vert ^{2}\tilde{\omega}_{3}\left( \tilde{\omega}_{1}+%
\tilde{\omega}_{2}\right) -\frac{\kappa }{2}\tilde{\omega}_{1}\tilde{\omega}%
_{2}\tilde{\omega}_{3}.  \label{21}
\end{eqnarray}%
Based on the relationship of eigenvectors and the initial conditions, the
solutions of field amplitudes read
\begin{eqnarray}
\tilde{A} &=&\sum_{j=1}^{4}D_{j}e^{-i\lambda _{j}t},  \label{22} \\
\tilde{C}_{j} &=&\sum_{j=1}^{4}\frac{i\chi _{j}}{\tilde{\omega}_{j}+i\lambda
_{j}}D_{j}e^{-i\lambda _{j}t},  \label{23}
\end{eqnarray}%
where%
\begin{eqnarray}
D_{1} &=&\frac{\left( \lambda _{2}-i\tilde{\omega}_{3}\right) \left(
i\lambda _{3}+\tilde{\omega}_{3}\right) \left( i\lambda _{4}+\tilde{\omega}%
_{3}\right) \prod \limits_{j}\left( \lambda _{1}-i\tilde{\omega}_{j}\right) }{%
g\left( \tilde{\omega}_{1}-\tilde{\omega}_{3}\right) \left( \tilde{\omega}%
_{3}-\tilde{\omega}_{2}\right) \prod \limits_{l\neq 1}\left( \lambda
_{1}-\lambda _{l}\right) },  \notag \\
D_{j\neq 1} &=&\frac{\left( \lambda _{j}-i\tilde{\omega}_{1}\right) \left(
\lambda _{j}-i\tilde{\omega}_{2}\right) \prod \limits_{l}\left( i\lambda _{l}+%
\tilde{\omega}_{3}\right) }{g\left( \tilde{\omega}_{3}-\tilde{\omega}%
_{1}\right) \left( \tilde{\omega}_{3}-\tilde{\omega}_{2}\right)
\prod \limits_{l\neq j}\left( \lambda _{j}-\lambda _{l}\right) }.  \label{24}
\end{eqnarray}%
Based on Eqs. (\ref{22},\ref{23}), the evolution dynamics of the populations
of the emitters and the cavity are $P_{j}=\left \langle \sigma _{j}^{\dag
}\sigma _{j}\right \rangle =\left \vert \tilde{C}_{j}\right \vert ^{2}$ and $%
P_{c}=\left \langle a^{\dag }a\right \rangle =\left \vert \tilde{A}\right \vert
^{2}$, respectively. For convenience, in the following discussion, we only
consider the case where the two auxiliary emitters have red and blue
detunings from the cavity mode with equal magnitudes, i.e., $\delta
_{1}=\delta ,$ and $\delta _{2}=-\delta .$ In this case, the central
frequency of the central peak of the hybrid system is resonant with the
cavity mode.

It can be clearly seen from Figs. \ref{FS2} as follows: (i) under different values of $\gamma$, the introduction of auxiliary emitters can effectively prolong the lifetime of the excited state of the target emitter; (ii) smaller $\gamma$ is more pronounced the strong coupling effect between the target emitter and the hybrid cavity system, as manifested in the prolongation of the excited‑state lifetime and the increase in the amplitude of Rabi oscillations; (iii) when $\gamma$ is small, a stronger strong coupling effect corresponds to a narrower linewidth of the central peak of the hybrid cavity, as exemplified by the correspondence between the red solid lines in Figs. \hyperref[FS1]{\ref{FS1}(d-f)} and those in Figs. \hyperref[FS2]{\ref{FS2}(d-f)}.


\begin{figure}[tbh]
\centering
\includegraphics[width=0.95\columnwidth]{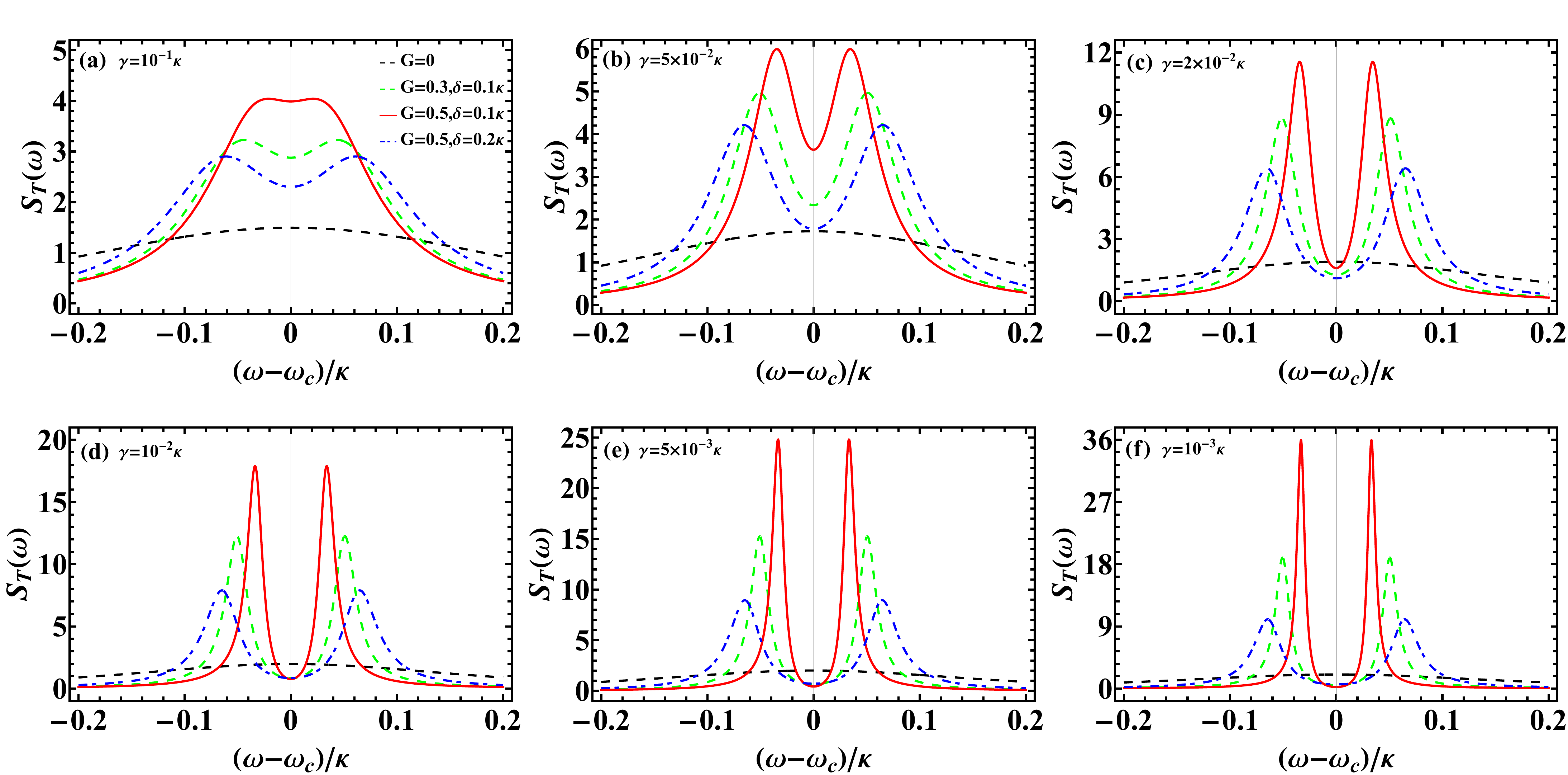}
\caption{Spontaneous emission spectrum $S_{T}(\omega)$ of the target emitter for different values of $\gamma$, $G$ and $\delta$.  (a) $\gamma=0.1\kappa$; (b) $\gamma=0.05\kappa$; (c) $\gamma=0.02\kappa$; (d) $\gamma=0.01\kappa$; (e) $\gamma=0.005\kappa$; (f) $\gamma=0.001\kappa$. Other parameters: $g = 0.25\kappa$ and $\delta_T = 0$.} \label{FS3}
\end{figure}

Next, we further verify the generation of the strong coupling effect of the
system by calculating the spontaneous emission spectrum of the target
emitter. Employing Eq. (\ref{23}) and the Fourier transform, we can obtain the
normalized spontaneous emission spectrum of the target emitter, i.e.,%
\begin{eqnarray}
S_{T}\left( \omega \right)  &=&\frac{s_{T}\left( \omega \right) }{\gamma_T
\int dt\left \langle \sigma _{T}^{\dag }\left( t\right) \sigma _{T}\left(
t\right) \right \rangle }  \notag \\
&=&\frac{s_{T}\left( \tilde{\omega}\right) }{\int d\omega s_{T}\left( \omega
\right) },  \label{25}
\end{eqnarray}%
where the power spectrum reads%
\begin{eqnarray}
s_{T}\left( \omega \right)  &=&\frac{\gamma_{T}}{2\pi}\int dt_{1}\int
dte^{i\omega \left( t-t_{1}\right) }\left \langle \sigma _{T}^{\dag }\left(
t_{1}\right) \sigma _{T}\left( t\right) \right \rangle   \notag \\
&=&\frac{\gamma_T}{2\pi }\left \vert \sum_{j=1}^{4}\frac{igD_{j}}{\tilde{\omega%
}_{3}+i\lambda _{j}}\int dte^{-i\left( \lambda _{j}+\omega _{c}\right)
t}e^{i\omega t}\right \vert ^{2}  \notag \\
&=&\frac{\gamma_T}{2\pi }\left \vert \sum_{j=1}^{4}\frac{gD_{j}}{\left( \tilde{%
\omega}_{3}+i\lambda _{j}\right) \left( \lambda _{j}-\omega +\omega
_{c}\right) }\right \vert ^{2},  \label{26}
\end{eqnarray}%
which can be derived by the quantum regression theorem \cite{57-Pelton_Storm_Leng_2019,58-Huang_Wang_Liang_Yu_2021}. As shown in Figs. \ref{FS3}, under resonance conditions $\delta_{T} = 0$, the spontaneous emission spectrum of the target emitter splits from a single peak into a symmetrical double peak. This further verifies that there is a strong coupling between the target emitter and the hybrid system. In addition, the larger the splitting interval, the shorter the period of Rabi oscillations in the time domain, the narrower the linewidth of the splitting peaks, and the longer the excited‑state lifetime in the time domain.


\end{document}